\documentclass[resetfootnote]{aastex701}

\usepackage{xcolor}

\begin{document}

\title{Magnetically Structured Oscillatory Power Along an Active-Region Transect in Near-UV {\sc Sunrise-iii}/SUSI Spectroscopy}

\author[orcid=0000-0002-7711-5397]{Shahin Jafarzadeh}
\affiliation{Astrophysics Research Centre, School of Mathematics and Physics, Queen’s University Belfast, Belfast, BT7 1NN, UK}
\affiliation{Max Planck Institute for Solar System Research, Justus-von-Liebig-Weg 3, 37077 G\"{o}ttingen, Germany}
\email[show]{shahin.jafarzadeh@qub.ac.uk}  

\author[orcid=0000-0002-9155-8039]{David B. Jess} 
\affiliation{Astrophysics Research Centre, School of Mathematics and Physics, Queen’s University Belfast, Belfast, BT7 1NN, UK}
\affiliation{Department of Physics and Astronomy, California State University Northridge, Northridge, CA 91330, USA}
\email[]{}  

\author[orcid=0000-0002-5365-7546]{Marco Stangalini} 
\affiliation{ASI Italian Space Agency, Via del Politecnico snc, I-00133 Rome, Italy}
\email[]{}  

\author[orcid=0000-0001-8556-470X]{Peter H. Keys}
\affiliation{Astrophysics Research Centre, School of Mathematics and Physics, Queen’s University Belfast, Belfast, BT7 1NN, UK}
\email[]{}  

\author[orcid=0000-0001-5170-9747]{Samuel D.~T. Grant} 
\affiliation{Astrophysics Research Centre, School of Mathematics and Physics, Queen’s University Belfast, Belfast, BT7 1NN, UK}
\email[]{}  

\author[orcid=0000-0003-3306-4978]{Timothy J. Duckenfield}
\affiliation{Astrophysics Research Centre, School of Mathematics and Physics, Queen’s University Belfast, Belfast, BT7 1NN, UK}
\email[]{} 

\author[orcid=0009-0003-6061-2404]{Glen Chambers}
\affiliation{Astrophysics Research Centre, School of Mathematics and Physics, Queen’s University Belfast, Belfast, BT7 1NN, UK}
\email[]{}  

\author[orcid=0000-0002-3418-8449,sname='Solanki']{Sami~K.~Solanki} \affiliation{Max Planck Institute for Solar System Research, Justus-von-Liebig-Weg 3, 37077 G\"{o}ttingen, Germany}\email{solanki@mps.mpg.de}	

\author[orcid=0000-0003-3490-6532,sname='Smitha']{H.~N.~Smitha} \affiliation{Max Planck Institute for Solar System Research, Justus-von-Liebig-Weg 3, 37077 G\"{o}ttingen, Germany}\email{narayanamurthy@mps.mpg.de}	

\author[orcid=0000-0003-1459-7074,sname='Lagg']{Andreas~Lagg} \affiliation{Max Planck Institute for Solar System Research, Justus-von-Liebig-Weg 3, 37077 G\"{o}ttingen, Germany}\email{lagg@mps.mpg.de}	

\author[orcid=0000-0002-9972-9840,sname='Gandorfer']{Achim~Gandorfer} \affiliation{Max Planck Institute for Solar System Research, Justus-von-Liebig-Weg 3, 37077 G\"{o}ttingen, Germany}\email{gandorfer@mps.mpg.de}

\author[orcid=0009-0009-4425-599X,sname='Feller']{Alex~Feller} \affiliation{Max Planck Institute for Solar System Research, Justus-von-Liebig-Weg 3, 37077 G\"{o}ttingen, Germany}\email{feller@mps.mpg.de}	

\author[orcid=0000-0003-1409-1145,sname='Iglesias']{Francisco~A.~Iglesias} \affiliation{Max Planck Institute for Solar System Research, Justus-von-Liebig-Weg 3, 37077 G\"{o}ttingen, Germany}\affiliation{Grupo de Estudios en Heliofísica de Mendoza, CONICET, Universidad de Mendoza, Boulogne sur Mer 683, 5500 Mendoza, Argentina}\email{iglesias@mps.mpg.de}	

\author[orcid=0000-0001-6317-4380,sname='Riethmüller']{Tino~L.~Riethmüller} \affiliation{Max Planck Institute for Solar System Research, Justus-von-Liebig-Weg 3, 37077 G\"{o}ttingen, Germany}\email{riethmueller@mps.mpg.de}	

\author[sname='Grauf']{Bianca~Grauf} \affiliation{Max Planck Institute for Solar System Research, Justus-von-Liebig-Weg 3, 37077 G\"{o}ttingen, Germany}\email{grauf@mps.mpg.de}	

\author[orcid=0000-0001-6029-7529,sname='Hoelken']{Johannes~Hoelken} \affiliation{Max Planck Institute for Solar System Research, Justus-von-Liebig-Weg 3, 37077 G\"{o}ttingen, Germany}\email{hoelken@mps.mpg.de}
\author[orcid=0000-0002-5054-8782,sname='Katsukawa']{Yukio~Katsukawa} \affiliation{National Astronomical Observatory of Japan, 2-21-1 Osawa, Mitaka, Tokyo 181-8588, Japan}\affiliation{Department of Astronomy, The University of Tokyo, 7-3-1, Hongo, Bunkyo-ku, Tokyo 113-0033, Japan}\affiliation{Department of Astronomical Science, The Graduate University for Advanced Studies (SOKENDAI), 2-21-1 Osawa, Mitaka, Tokyo 1818588, Japan}\email{yukio.katsukawa@nao.ac.jp}	
\author[orcid=0000-0002-0787-8954,sname='Bernasconi']{Pietro~Bernasconi} \affiliation{Johns Hopkins University Applied Physics Laboratory, 11100 Johns Hopkins Road, Laurel, Maryland, USA}\email{pietro.bernasconi@jhuapl.edu}	
\author[sname='Berkefeld']{Thomas~Berkefeld} \affiliation{Institut für Sonnenphysik (KIS), Georges-Köhler-Allee 401a, 79110 Freiburg, Germany}\email{thomas.berkefeld@leibniz-kis.de}	
		
\author[orcid=0000-0001-9228-3412,sname='Álvarez-Herrero']{Alberto~Álvarez-Herrero} \affiliation{Instituto Nacional de T\'ecnica Aeroespacial (INTA), Ctra. de Ajalvir, km. 4, E-28850 Torrejón de Ardoz, Spain}\affiliation{Spanish Space Solar Physics Consortium}\email{alvareza@inta.es}	
\author[orcid=0000-0001-5616-2808,sname='Kubo']{Masahito~Kubo} \affiliation{National Astronomical Observatory of Japan, 2-21-1 Osawa, Mitaka, Tokyo 181-8588, Japan}\email{masahito.kubo@nao.ac.jp}	
\author[orcid=0000-0001-8829-1938,sname='Orozco~Suárez']{David~Orozco~Suárez} \affiliation{Instituto de Astrofísica de Andalucía, CSIC, Glorieta de la Astronomía s/n, 18008 Granada, Spain}\affiliation{Spanish Space Solar Physics Consortium}\email{orozco@iaa.es}	
\author[sname='Carpenter']{Michael~Carpenter} \affiliation{Johns Hopkins University Applied Physics Laboratory, 11100 Johns Hopkins Road, Laurel, Maryland, USA}\email{michael.carpenter@jhuapl.edu}	
\author[sname='Bell']{Alexander~Bell} \affiliation{Institut für Sonnenphysik (KIS), Georges-Köhler-Allee 401a, 79110 Freiburg, Germany}\email{albe@leibniz-kis.de}	
\author[orcid=0000-0001-7764-6895,sname='Martínez~Pillet']{Valentín~Martínez~Pillet} \affiliation{Instituto de Astrofísica de Canarias, Vía Láctea, s/n, E-38205 La Laguna, Spain}\affiliation{Spanish Space Solar Physics Consortium}\email{vmpillet@iac.es}
		
\author[orcid=0000-0002-7318-3536,sname='Bailén']{Francisco~Javier~Bailén} \affiliation{Instituto de Astrofísica de Andalucía, CSIC, Glorieta de la Astronomía s/n, 18008 Granada, Spain}\affiliation{Spanish Space Solar Physics Consortium}\email{fbailen@iaa.es}	
\author[orcid=0000-0002-2055-441X,sname='Blanco~Rodríguez']{Julian~Blanco~Rodríguez} \affiliation{Universitat de Valencia Catedrático José Beltrán 2, E-46980 Paterna-Valencia, Spain}\affiliation{Spanish Space Solar Physics Consortium}\email{julian.blanco@uv.es}	
\author[orcid=0000-0003-4319-2009,sname='Castellanos~Durán']{Juan~Sebastián~Castellanos~Durán} \affiliation{Max Planck Institute for Solar System Research, Justus-von-Liebig-Weg 3, 37077 G\"{o}ttingen, Germany}\email{castellanos@mps.mpg.de}	
\author[orcid=0009-0002-6808-5154,sname='Harnes']{Edvarda~Harnes} \affiliation{Max Planck Institute for Solar System Research, Justus-von-Liebig-Weg 3, 37077 G\"{o}ttingen, Germany}\email{harnes@mps.mpg.de}	
\author[orcid=0000-0002-4669-5376,sname='Ishikawa']{Ryohtaroh~T.~Ishikawa} \affiliation{National Institute for Fusion Science, 322-6 Oroshi-cho, Toki City 509-5292, Japan}\email{ishikawa.ryohtaro@nifs.ac.jp}	
\author[orcid=0000-0001-7452-0656,sname='Kawabata']{Yusuke~Kawabata} \affiliation{National Astronomical Observatory of Japan, 2-21-1 Osawa, Mitaka, Tokyo 181-8588, Japan}\email{kawabata.yusuke@nao.ac.jp}	
\author[orcid=0000-0002-1043-9944,sname='Matsumoto']{Takuma~Matsumoto} \affiliation{Centre for Integrated Data Science, Institute for Space-Earth Environmental Research, Nagoya University, Furocho, Chikusa-ku, Nagoya, Aichi 464-8601, Japan}\email{takuma.matsumoto@gmail.com}	
\author[orcid=0000-0002-7044-6281,sname='Oba']{Takayoshi~Oba} \affiliation{Advanced Research Center for Space Science and Technology, Institute of Science and Engineering, Kanazawa University, Kakuma-machi, Kanazawa, Ishikawa 920-1192, Japan}\affiliation{Max Planck Institute for Solar System Research, Justus-von-Liebig-Weg 3, 37077 G\"{o}ttingen, Germany}\email{oba@mps.mpg.de}	
\author[orcid=0000-0003-0175-6232,sname='Siu-Tapia']{Azaymi~L.~Siu-Tapia} \affiliation{Instituto de Astrofísica de Andalucía, CSIC, Glorieta de la Astronomía s/n, 18008 Granada, Spain}\affiliation{Spanish Space Solar Physics Consortium}\email{siu@iaa.es}	
\author[orcid=0000-0003-1483-4535,sname='Strecker']{Hanna~Strecker} \affiliation{Instituto de Astrofísica de Andalucía, CSIC, Glorieta de la Astronomía s/n, 18008 Granada, Spain}\affiliation{Spanish Space Solar Physics Consortium}\email{streckerh@iaa.es}	
\author[orcid=0000-0003-1971-5551,sname='Vukadinović']{Dušan~Vukadinović} \affiliation{Institut für Physik, Universität Graz, Universitätsplatz 5, 8010 Graz, Austria}\affiliation{Max Planck Institute for Solar System Research, Justus-von-Liebig-Weg 3, 37077 G\"{o}ttingen, Germany}\email{vukadinovic@mps.mpg.de}	

\begin{abstract}
We present a multi-line characterisation of how oscillatory power is organised across distinct magnetic environments in an active region using seeing-free, stratospheric near-ultraviolet spectroscopy from the \textsc{Sunrise-III} UV Spectropolarimeter and Imager (SUSI). A two-hour time series of short raster scans in the line-rich 327--329\,nm window samples along a single transect that contains the following regions: weak magnetic-field surroundings, a plage, a sunspot, and a pore. From a set of 30 selected, relatively unblended absorption lines, we extract line-core Doppler-velocity time series and compute Morlet-wavelet refined global spectra from which we form band-integrated power maps for three frequency bands (2--4, 4--6, and 6--12\,mHz). The stacked, line-resolved maps reveal a clear environment-dependent redistribution of power: 2--4\,mHz power is strongest in the weak-field/plage segments but is commonly suppressed in the umbra and pore cores, while 4--6\,mHz and 6--12\,mHz power becomes relatively enhanced in the strongest-field regions, with line-dependent behaviour in the penumbra and plage. Across the line ensemble, this broad frequency structuring is coherent, but the detailed spatial distribution and relative band ranking are not identical from line to line -- even among spectral lines with comparable effective formation depths -- demonstrating clear line dependence. This novel result implies that single-line measurements may miss secondary components of the local wave spectrum because different lines weight co-existing perturbations and modes differently; therefore, the SUSI near-UV window provides a uniquely diagnostic-rich mapping of oscillations, offering leverage that is difficult to obtain with traditional one- or two-line approaches.
\end{abstract}

\section{Introduction}
\label{sec:introduction}

Waves and oscillations in the lower solar atmosphere are ubiquitous, but their power distributions and dominant periodicities vary strongly with local atmospheric and magnetic conditions \citep{2023LRSP...20....1J}. In and around active regions, oscillatory signals reorganise on small spatial scales as the field strength, inclination, and thermodynamic stratification change, producing distinct behaviour in sunspot umbrae, penumbrae, pores, plage/network, and adjacent weak-field areas \citep[e.g.,][]{2006RSPTA.364..313B, 2012RSPTA.370.3193D, 2015LRSP...12....6K, 2015SSRv..190..103J}. This diversity is not merely a change in amplitude: characteristic frequency content, intermittency, and lateral spatial coherence can differ between environments, reflecting both wave/oscillation physics (including magnetic filtering and height-dependent transmission in structured atmospheres) and the visibility of different perturbations in spectral-line diagnostics \citep{2016ApJ...828...23S}.

A classical observational contrast is between umbral and penumbral regimes. Umbral spectra often show prominent power near the three-minute and five-minute ranges, whereas penumbral oscillations can exhibit different dominant periods and spatial patterns, including running penumbral waves that trace outward-propagating magnetoacoustic waves in chromospheric diagnostics \citep[e.g.,][]{1998ApJ...497..464L, 2002A&A...387.1092J, 2003A&A...403..277R, 2006ApJ...640.1153C, 2007ApJ...671.1005B, 2010ApJ...722..131F, 2013ApJ...779..168J, 2015A&A...580A..53L}. These differences are commonly interpreted in terms of acoustic cut-off modification by magnetic inclination, wave guiding, and mode conversion \citep{1995ApJ...444..879W, 1977A&A....55..239B, 2011SoPh..268..349S}. Pores can also show distinct frequency distributions depending on size, topology, and thermal structure \citep{2015A&A...579A..73M, 2018ApJ...857...28K, 2021A&A...649A.169S, 2022ApJ...938..143G}, and recent analyses indicate that power spectra of both sunspots and pores can be multi-component, with coexisting peaks suggesting multiple magnetohydrodynamic (MHD) wave modes \citep[e.g.,][]{2014A&A...569A..72S, 2023A&A...674A.109C}. Higher-order eigenmodes have also been identified in both sunspots and pores \citep{2017ApJ...842...59J, 2022ApJ...927..201A, 2022NatCo..13..479S, 2024A&A...688A...2J}. In plage and network, power is reshaped by local magnetic geometry, producing localised enhancements and suppressions relative to nearby quiet Sun \citep{1992ApJ...393..782T, 2013SoPh..287..107R, 2016ApJ...828...23S}. Even in internetwork/weak-field areas, $p$-mode oscillations dominate much of the atmospheric signal, while small-scale magnetic elements can imprint localised modifications to both power and phase behaviour \citep{2002RvMP...74.1073C, 2013ApJ...779..168J}. Overall, modern time-frequency analyses increasingly support the view that spectra in magnetic structures are frequently multi-component, with non-negligible power beyond a single dominant band \citep{2022NatCo..13..479S}.

Most observational studies, however, rely on one or two spectral lines (or a small set of observables), often chosen to represent a few well-separated layers. While these works have yielded major insights, they may conflate environment-dependent dynamics with line-dependent response: (i) the effective formation depth of a line can shift laterally as physical conditions change, even across small spatial scales \citep{2023A&A...669A.144S}; and (ii) even diagnostics sampling of similar nominal heights can weight different perturbations (velocity, temperature, opacity/source-function variations) differently, so that the observed oscillation signatures depend on the chosen line as well as the underlying dynamics \citep[e.g.,][]{1992ApJ...398..375R, 2008A&A...480..515C, 2012ApJ...749..136L}. Multi-line spectroscopy offers a direct route past these limitations by providing many simultaneous observables with diverse radiative sensitivities and overlapping (but not identical) height weighting.

The near-ultraviolet (near-UV) spectral region is particularly attractive because it contains a dense forest of lines from multiple species \citep{1973apds.book.....D}, but it has historically been challenging for ground-based wave studies. The third flight of the \textsc{Sunrise} balloon-borne solar observatory \citep{2025SoPh..300...75K, solankietal2026} has opened a new regime: seeing-free, high-cadence near-UV spectroscopy suitable for time-frequency analysis across many lines at once, thus extending considerably the capabilities of earlier versions of \textsc{Sunrise} \citep{2010ApJ...723L.127S, 2011SoPh..268....1B, 2017ApJS..229....2S}. In \citet[][hereafter referred to as Paper~I]{Jafarzadeh+2026_PaperI} we used this capability to examine multi-line frequency structuring within the sunspot umbra; here we shift emphasis to the lateral organisation of oscillatory power along a magnetically diverse transect. We analyse a two-hour \textsc{Sunrise-III}/SUSI time series in the 327--329\,nm spectral window in which the slit samples, within the same scan geometry, weak-field surroundings, plage, a sunspot (umbra and penumbra), and a nearby pore. From a set of relatively unblended atomic lines (excluding molecular features), we extract line-core Doppler velocities and compute refined global wavelet spectra for each slit position and line. We then map band-integrated power in three frequency intervals (2--4, 4--6, and 6--12\,mHz) and compare the resulting patterns across the full line ensemble, treating the diagnostics primarily in a line-resolved sense.

The aim of this Letter is therefore twofold: (i) to provide a compact multi-line characterisation of how low-, mid-, and high-frequency power redistributes along a sunspot--plage--pore transect in the near-UV; and (ii) to demonstrate explicitly how strongly the inferred frequency organisation depends on the chosen spectral line, even within a narrow spectral window, consistent with line-dependent diagnostic response and environment-dependent mixtures of oscillatory phenomena.

\section{Data}
\label{sec:data}

The data analysed in this Letter were obtained with the \textsc{Sunrise-III} UV Spectropolarimeter and Imager (SUSI; \citealt{2025SoPh..300...65F, 2025SoPh..300...58I}) during the \textsc{Sunrise} balloon flight on 14~July~2024, between 21:40 and 23:39~UTC. The target is a large active region observed close to disc centre ($\mu\!\approx\!0.96$), including a sunspot with umbra and penumbra, a nearby pore, and surrounding plage and quieter regions.

SUSI was operated in a small-raster mode in the 327--329\,nm window. The slit spans $\sim$42\arcsec\ along its length and the raster covers a narrow strip of width $\sim$3\arcsec\ in the scan direction, with the slit oriented $\sim$24$^{\circ}$ from the north--south direction. Owing to the stable gondola pointing \citep{2025SoPh..300..112B} and the Correlation Wavefront Sensor (CWS; \citealt{2026arXiv260207448B}), the slit position remained stable throughout the sequence. One raster cycle was completed every $\sim$39\,s, yielding a two-hour time series of 180 scans and enabling wave diagnostics up to $\sim$13\,mHz. A short calibration-related interruption introduces a data gap; for the time-series analysis, the missing samples are filled by linear interpolation.

In this programme, SUSI recorded full-Stokes spectra in the selected near-UV window, but at the time of the present early-science analysis the fully reduced/calibrated polarimetric products were not yet available. The present Letter is therefore restricted to Stokes~$I$ only, with the polarimetric information reserved for follow-up work once the relevant products become available. The data were processed with the standard SUSI reduction pipeline, including dark and flat-field corrections and wavelength calibration \citep[for more information, see][]{solankietal2026}. In the present Letter, the phase-diversity-reconstructed slit-jaw images are used only for contextual visualisation and slit-location reference, whereas the quantitative wave analysis is performed directly on the reduced spectral rasters themselves. In additional post-processing, time-series raster images were assembled from the individual slit positions and further corrected for solar rotation, field derotation, and residual image jitter, resulting in a four-dimensional data cube with axes $(t,\,x,\,y,\,\lambda)$. The raster images have a spatial sampling of $0{\,}.{\!\!}''03$~px$^{-1}$, and the spectral sampling is $\sim$8.3~m\AA\,px$^{-1}$ in this window.

The 327--329\,nm range contains a dense set of photospheric and low-chromospheric absorption lines (see Paper~I for the full set of line identifications; see also Appendix~\ref{sec:appendix_eb} for a summary view of the full window, with the analysed subset highlighted). For contextual guidance only (not as a primary interpretation resource), we provide approximate formation-height estimates using the Eddington--Barbier (EB) approximation, i.e., the geometric height where $\tau_{\lambda}=1$ for each wavelength sample in representative model atmospheres; these EB heights are summarised in Appendix~\ref{sec:appendix_eb}.

For the present study, we analyse the middle slit position (a cut through the centre of the raster field-of-view) as a representative transect crossing weak-field surroundings, plage, middle of the sunspot (umbra/penumbra) and the pore. To improve robustness in low-signal regions and suppress pixel-scale residuals, we apply a $3\times3$ spatial boxcar average prior to extracting the transect, which reduces the dominant uncorrelated component of the pixel-scale noise by a factor of $\sqrt{9}=3$ and corresponds to an additional blur of order $\sim0{\,}.{\!\!}''08$; this is comparable to the 1-m diffraction limit at 328\,nm ($\sim0{\,}.{\!\!}''083$). In addition, prior to line-core fitting we apply a mild 1D spectral smoothing using a boxcar of width 3 pixels (equivalent to a small triangular kernel), targeting high-frequency spectral noise on sub-resolution scales. With a sampling of $\sim$0.83~pm\,px$^{-1}$ and an effective spectral resolution of $\sim$2.39~pm, the 3-point boxcar corresponds to an equivalent broadening of $\mathrm{FWHM}_{\mathrm{eq}}\!\approx\!1.92$~px ($\approx$1.59~pm), i.e., well below the instrumental resolution, and is intended to reduce high-frequency spectral noise without materially altering resolvable line-core structure.
These smoothing steps are intentionally mild and local, and are used only as preprocessing measures to improve the stability of the line-centre fitting.

For the wave analysis, we use line-core Doppler-velocity time series derived for each position along the transect and each time step, using an adaptive Voigt-fitting procedure that identifies the absorption minimum near each rest wavelength and fits the local profile to obtain the line-centre shift \citep{2026FrASS...Jafarzadeh_inprep}. The resulting line-of-sight (LOS) velocities, $v_{\mathrm{LOS}}(s,t;\lambda_j)$, with $s$ denoting position along the slit, are extracted for a set of 30 relatively unblended atomic lines and form the basis for all wave-power diagnostics presented below. Representative fitted LOS velocities in the analysed transect are typically in the range $-1.55${\,}--{\,}$2.04$~km\,s$^{-1}$ (central 99\% of values; outliers excluded), with the exact spread depending on spectral line and solar environment. For the observational data analysed here, the true line-centre position is not known, so the true per-point absolute centre error cannot be measured directly from the SUSI spectra themselves. The fitting methodology has, however, been benchmarked separately against synthetic spectra with known ground truth, where centre-recovery errors are quantified across lines of different complexity \citep{2026FrASS...Jafarzadeh_inprep}. Thus, while small non-persistent fitting perturbations may occur in the extracted time series, they are not expected to dominate the statistically significant oscillatory signals on which the present analysis is based. Because both atmospheric conditions and spectral-line response vary strongly along the transect, the results are presented primarily in a line-resolved sense; further fitting details are described in Paper~I.

\section{Analysis and Results}
\label{sec:analysis}

\begin{figure}[!ht]
\centering
\includegraphics[width=\linewidth]{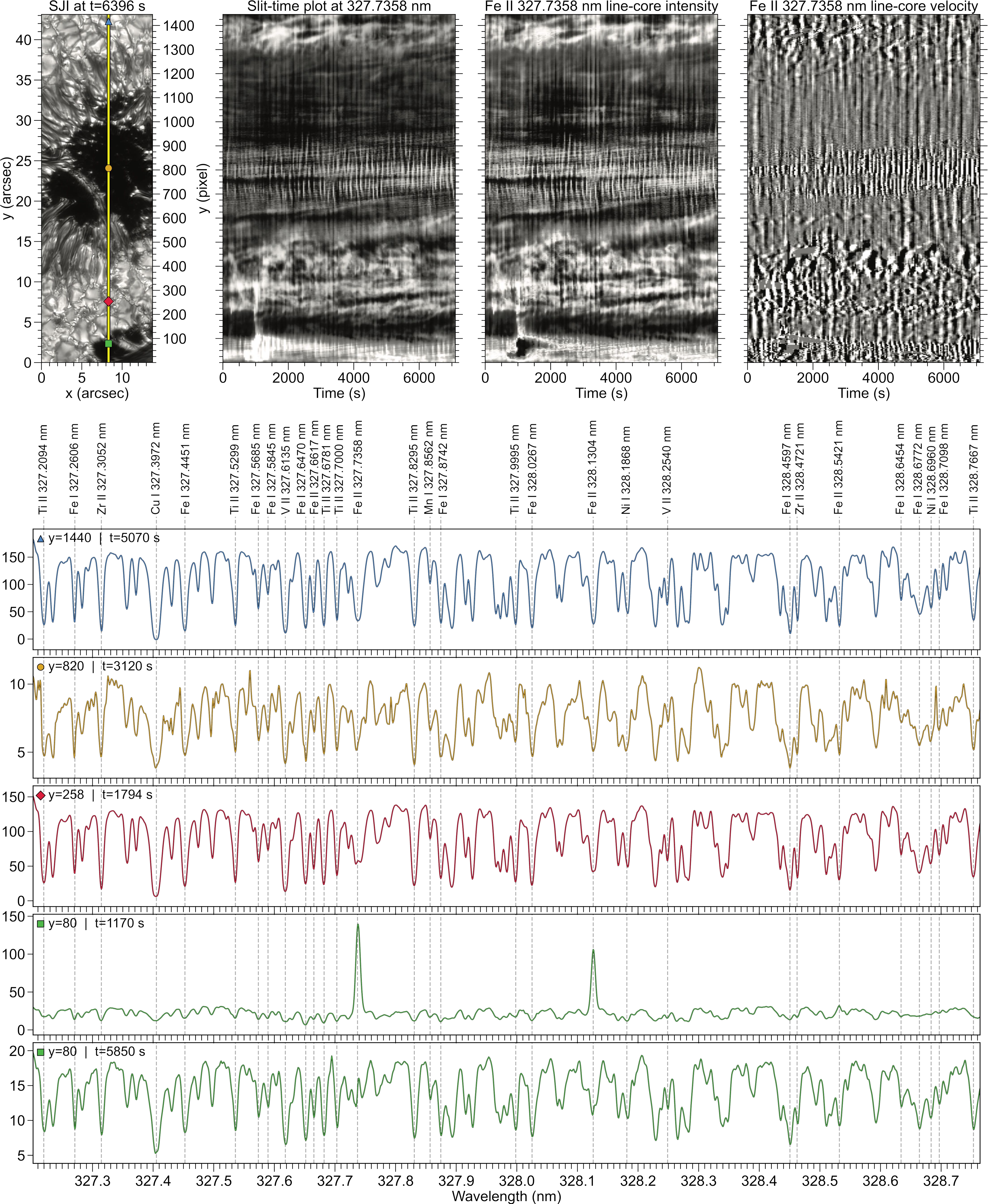}
\caption{Context for the active-region transect and spectral variability along the SUSI slit. 
Top-left: slit-jaw image (SJI) with the analysed transect overplotted; coloured markers indicate the locations used for the example spectra. 
Remaining top-row panels (left to right): slit--time diagrams for Fe\,\textsc{ii} 327.7358\,nm showing (i) intensity at a fixed wavelength sample near the atlas wavelength, (ii) fitted line-core intensity, and (iii) fitted line-core LOS velocity, displayed as a visualisation of the LOS-velocity amplitude.
Bottom row: five example spectra at the marked locations, shown as relative intensity (arbitrary units) and lightly smoothed for readability (visualisation only). Vertical lines indicate the analysed spectral lines.}
\label{fig:context_figure}
\end{figure}

Figure~\ref{fig:context_figure} sets the observational context and motivates the line-resolved approach adopted in this Letter. It summarises the slit geometry and highlights the strong spatial and temporal variability of the near-UV spectra that underpins our line-by-line analysis. The top-left panel shows a representative slit-jaw image (SJI) with the fixed transect overplotted, together with four marked reference locations along the slit. The remaining panels in the top row show slit-time diagrams for a representative line (Fe\,\textsc{ii} 327.7358\,nm): (i) intensity at a fixed wavelength sample near the atlas rest wavelength, (ii) fitted line-core intensity (often labelled as `Doppler compensated intensity'), and (iii) fitted line-core LOS velocity. These panels demonstrate that the character of variability changes markedly along the slit, with faster, higher-contrast fluctuations in the umbra and pore compared to surrounding regions, and additional transient behaviour near the pore early in the time series.

The bottom block of Figure~\ref{fig:context_figure} shows five example spectra (lightly smoothed for Figure readability) extracted at the marked slit locations. From top to bottom: (1) representative weak-field photospheric plasma near the spot (sampling a granular location), showing predominantly absorption profiles across the window; (2) a location close to the sunspot centre, with a substantially reduced overall intensity level and noticeable profile-shape changes in several diagnostics; (3) a plage location, exhibiting systematic differences relative to the weak-field spectrum, including changes in line widths and small emission-like features in some lines (e.g., Fe\,\textsc{ii} 327.7358\,nm); (4--5) a location near the pore centre shown at two different times, illustrating the strong time dependence of the profiles there. In particular, the spectrum at $t\!\approx\!1170$\,s occurs during a transient brightening, during which two Fe\,\textsc{ii} lines (327.7358 and 328.1304\,nm) develop strong emission that compresses the apparent contrast of the surrounding absorption spectrum, whereas at a later time the same location returns to a more typical absorption-dominated appearance. We do not attempt a detailed physical interpretation of this brief brightening in the present Letter; its purpose here is simply to illustrate the strong time dependence that can occur locally in some spectral lines. The marked vertical dashed lines indicate the set of analysed atomic lines and highlight that profile morphology can vary substantially between environments and with time within the same dataset.

To aid visual inspection, the slit--time panels in Figure~\ref{fig:context_figure} are displayed with visualisation-only contrast adjustments (robust scaling and mild detrending) to make coherent patterns easier to see. All quantitative results presented below are computed from the fitted $v_{\mathrm{LOS}}(s,t;\lambda_j)$ time series following the procedures outlined in Section~\ref{subsec:wavelet_method}. Occasional transient excursions (such as the pore brightening) affect only a limited fraction of the two-hour sequence; the subsequent wavelet analysis further restricts the interpretation to statistically significant power, so the main band-integrated trends reported in this Letter are not driven by a small number of extreme frames.

To facilitate inspection of the full spatio-temporal spectral variability beyond the few examples shown in Figure~\ref{fig:context_figure}, we provide an interactive web viewer\footnote{\url{https://WaLSA.team/Sunrise/SUSI}} that allows the spectra to be explored as a function of slit position and time, with the analysed line positions overplotted.

\subsection{Wavelet power diagnostics}
\label{subsec:wavelet_method}

For each spectral line, $\lambda_j$, and slit position, $s$, we analyse the LOS-velocity time series $v_{\mathrm{LOS}}(s,t;\lambda_j)$ using Morlet wavelets. Each time series is linearly detrended, to remove slow background drifts unrelated to the oscillatory signal of interest, and apodised with a Tukey window (10\% taper) to reduce edge discontinuities and spectral leakage in the wavelet transform. We compute the wavelet power spectrum and retain only statistically significant power above the 95\% confidence level, excluding the cone of influence (CoI) to suppress edge effects. This significance filtering also helps suppress the influence of small, non-persistent perturbations in the fitted time series, so that the final wavelet-based spectrum emphasises coherent oscillatory power rather than isolated excursions. We then form the refined global wavelet spectrum (RGWS), i.e., the time-integrated significant power as a function of frequency, $P(f; s,\lambda_j)$, which compactly captures both dominant and weaker-but-significant components while reducing sensitivity to transient edge artefacts. In addition, we suppress very low-frequency components ($\nu \lesssim 2$\,mHz), which primarily reflect slow background evolution rather than the wave dynamics of interest here. The wavelet analysis is performed with the WaLSAtools\footnote{\url{https://github.com/WaLSAteam/WaLSAtools}} \citep{2025NRvMP...5...21J, walsatools..2025...17569951} implementation, following the same strategy as in Paper~I.

\subsection{Band-integrated power maps}
\label{subsec:slit_power_maps}

To visualise how oscillatory power redistributes along the slit across the full line ensemble, we integrate the RGWS within three frequency intervals (2--4\,mHz, 4--6\,mHz, and 6--12\,mHz) and construct 1D, along-slit band-integrated power maps. For each line, $\lambda_j$, the band-integrated power is,
\begin{equation}
P_{\mathrm{band}}(s;\lambda_j) \;=\; \int_{f_0}^{f_1} P(f; s,\lambda_j)\, \mathrm{d}f \ ,
\end{equation}
where $P(f; s,\lambda_j)$ contains only significant power outside the CoI, and $f_0$ and $f_1$ denote the lower and upper frequency bounds of the band being integrated. Because different spectral lines have different wave sensitivities and absolute velocity amplitudes, we do not interpret absolute power differences between lines. Instead, to highlight where along the slit each line shows its strongest power in a given band, we normalise each line by its maximum band power along the slit. This yields a line-by-line map that emphasises spatial organisation and frequency-dependent redistribution along the transect.
We adopt these three comparatively broad bands deliberately, in order to capture the principal low-, intermediate-, and high-frequency redistribution trends in a compact Letter-format analysis. A more detailed decomposition of the fine structure within, for example, the 3-min band is certainly of interest, but is deferred to follow-up work.

\begin{figure}[!t]
\centering
\includegraphics[width=\linewidth]{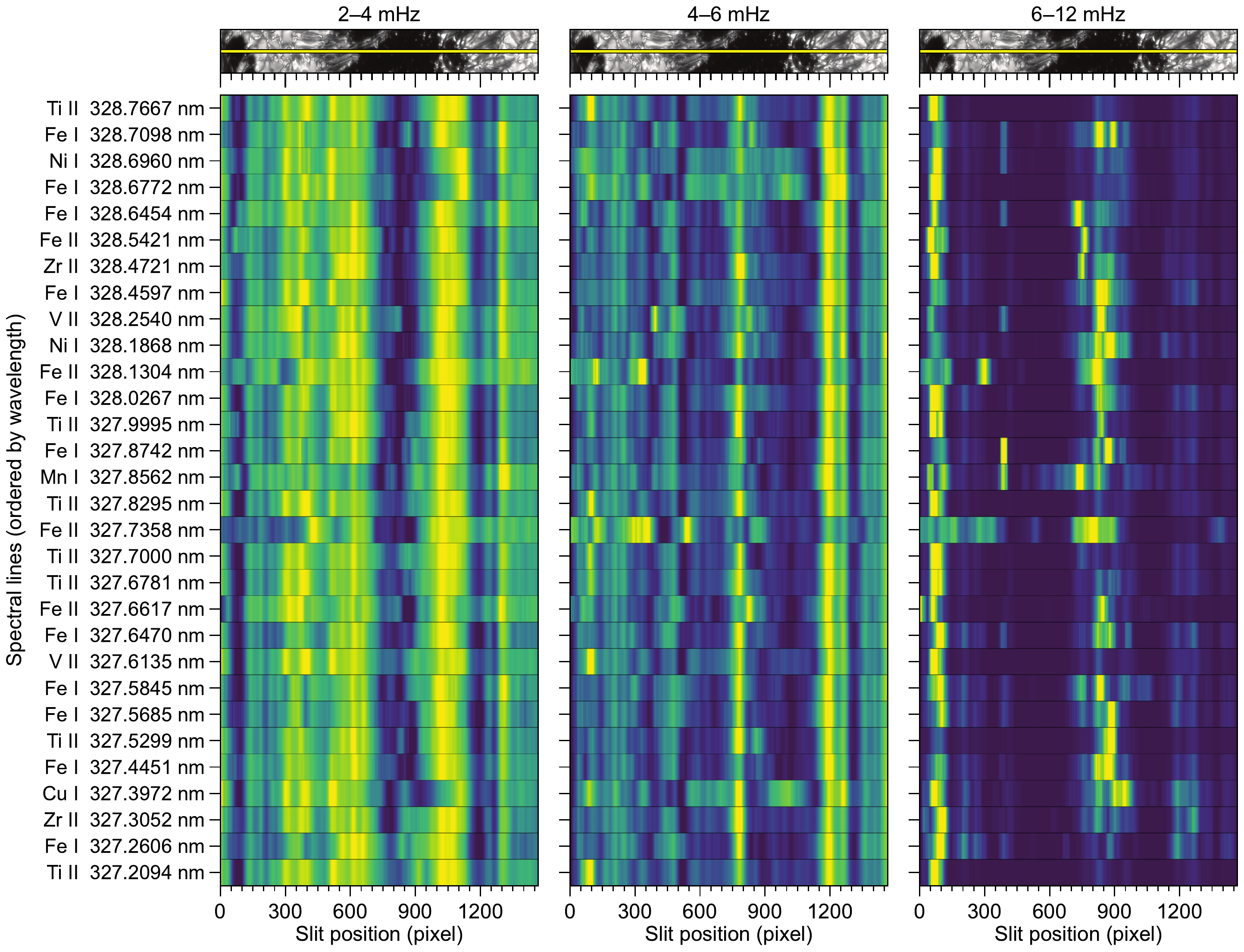}
\caption{Stacked, line-resolved band-integrated RGWS power along the slit for three frequency ranges (left to right): 2--4\,mHz, 4--6\,mHz, and 6--12\,mHz. Rows correspond to the 30 selected spectral-line diagnostics (ordered by increasing wavelength). Each row is normalised to its own maximum along the slit, so the colour scale highlights \emph{where} along the slit each line concentrates power within the given band rather than comparing absolute power between different lines. A slit-jaw context strip above each panel marks the slit position and facilitates identification of umbra, penumbra, plage, and pore segments.}
\label{fig:rgws_bandpass_maps}
\end{figure}

Figure~\ref{fig:rgws_bandpass_maps} shows the stacked band-power maps for the 30 analysed lines, ordered by increasing wavelength. A narrow SJI strip above each panel provides the spatial context and indicates where the slit samples umbra, penumbra, plage, and the pore. Three key trends emerge. First, the 2--4\,mHz band shows relative suppression in the strongest-field cores (umbra and pore centres) and also in a highly structured penumbral segment (around the slit position 1200\,pixel) along the slit, compared to surrounding regions. Second, the 4--6\,mHz band shows enhanced relative power in the umbral core and in the same penumbral segment for a large fraction of lines, with a more variable response in the pore and plage. Third, the 6--12\,mHz band is most prominently concentrated in the umbra and pore cores for most diagnostics, with occasional enhancements in plage in a smaller subset of lines. Importantly, the detailed spatial pattern is not identical across lines, underscoring that the inferred organisation depends on the chosen spectral line even within this narrow window. For completeness, Appendix~\ref{app:rgws_additional_figures} provides a supplementary Blue-Yellow-Red composite visualisation of the same three band-integrated maps, intended only as a qualitative overview; the quantitative interpretation in this Letter is based on Figures~\ref{fig:rgws_bandpass_maps} and \ref{fig:rgws_bandpass_stats}.

\subsection{Ensemble summary across the line set}
\label{subsec:rgws_bandpass_stats}

\begin{figure}[!t]
\centering
\includegraphics[width=\linewidth]{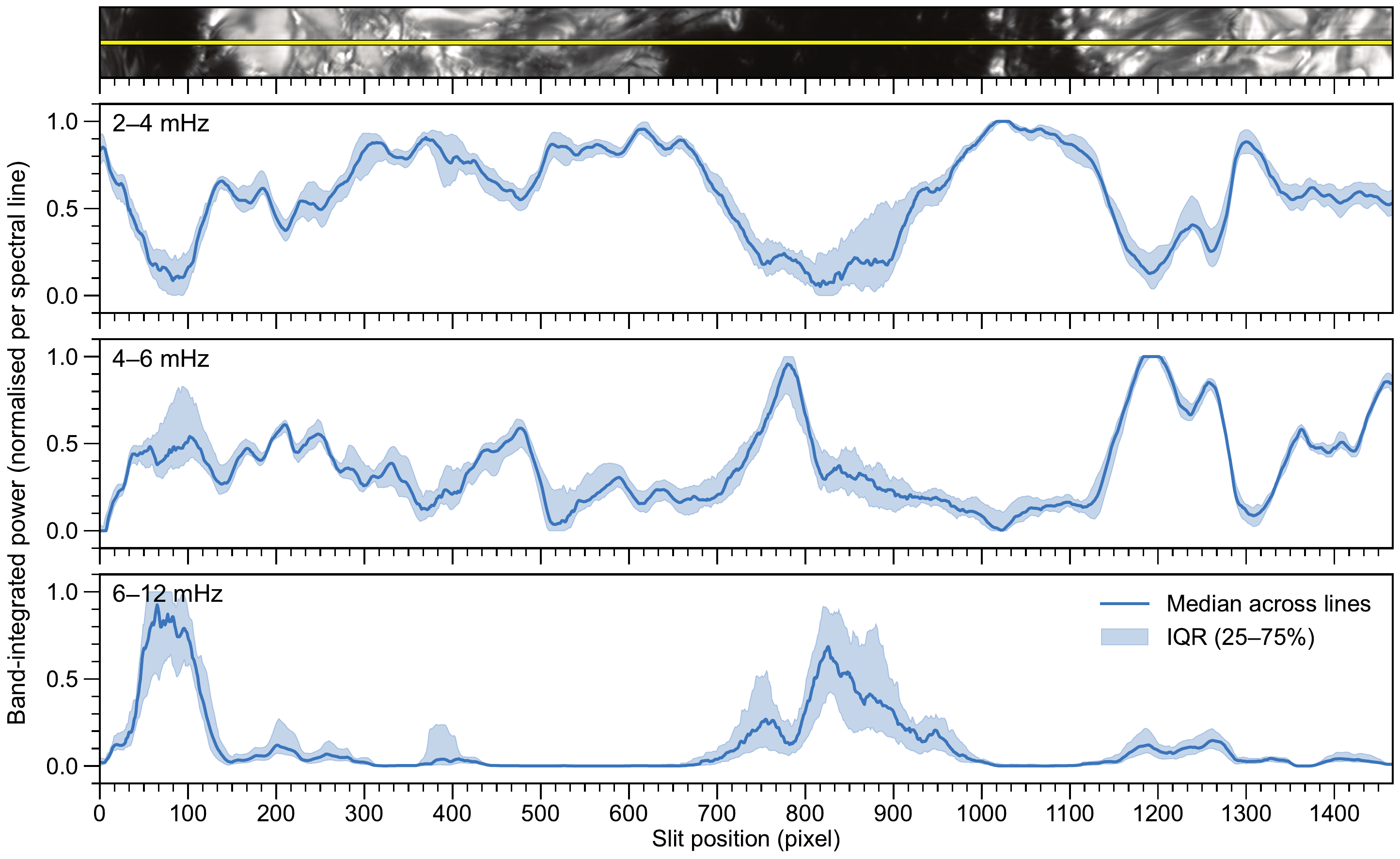}
\caption{Robust ensemble summary of the line-resolved band-integrated maps in Figure~\ref{fig:rgws_bandpass_maps}. For each frequency band, the solid curve shows the median of the row-normalised band power across the 30 lines as a function of slit position, and the shaded region shows the interquartile range (25th--75th percentile). The SJI strip above these panels provides the same spatial context as in Figure~\ref{fig:rgws_bandpass_maps}.}
\label{fig:rgws_bandpass_stats}
\end{figure}

While Figure~\ref{fig:rgws_bandpass_maps} highlights the line-to-line diversity, it is also useful to quantify how coherent the redistribution is across the ensemble of near-UV spectral lines. We therefore compute robust summary statistics across the set of row-normalised band-power profiles at each slit position. Figure~\ref{fig:rgws_bandpass_stats} shows the median profile across lines (solid curve) together with the interquartile range (IQR; shaded region) for each frequency band.

These ensemble profiles provide a compact ``collapse'' of the stacked maps and highlight where a majority of lines agree on enhanced (or suppressed) relative power. We emphasise that this ensemble summary is complementary to the stacked maps: because it aggregates spectral lines with different height-weighting and radiative sensitivities, line-specific (and potentially physically informative) localised features in individual spectra may be diluted in the median/IQR representation. Pronounced median enhancements indicate slit locations where many diagnostics simultaneously place their strongest relative power within that band, whereas a broadened IQR indicates increased line-to-line diversity, consistent with the diagnostic-dependent patterns seen in Figure~\ref{fig:rgws_bandpass_maps}. 

Figure~\ref{fig:rgws_bandpass_stats} further shows that the median 2--4\,mHz profile peaks primarily in the weak-field/plage parts of the transect, whereas enhanced 4--6\,mHz power concentrates in the umbra, the penumbral segment (around the slit position 1200\,pixel) and also partly in the pore, and enhanced 6--12\,mHz power is most localised within the strongest-field structures, with distinct peaks in the sunspot umbra and the pore. The IQR broadens most strongly in the umbra and pore, indicating increased line-to-line diversity where multi-component spectra and a mixture of MHD wave modes are commonly expected. This behaviour is physically plausible because the selected near-UV lines span a range of radiative sensitivities and (proxy) height weightings, so different spectral lines can respond differently to co-existing perturbations and local conditions within the same magnetic structure. In addition, because each line samples a finite vertical range rather than a single geometrical height, propagating disturbances with phase variations across that range can in principle undergo partial cancellation in the emergent Doppler signal, especially where the phase varies rapidly with height. This effect is part of the line-dependent response discussed here, but it does not introduce any additional cross-line cancellation in the construction of the line-resolved power maps themselves, since each spectral line is analysed independently. For context, EB proxy estimates for the 30 lines (Appendix~\ref{sec:appendix_eb}) place their characteristic $\tau_\lambda\!=\!1$ heights in the photosphere to low chromosphere. In weak-field and plage regions, they span roughly $\sim$26--800\,km with a median around 370--400\,km. In a representative umbral atmosphere, the corresponding heights are shifted downward by a few hundred kilometres because of the Wilson depression. These values are used only as qualitative guidance and do not imply a fixed height ordering along the full transect.

\section{Discussion and Conclusions}
\label{sec:discussion}

We have used seeing-free \textsc{Sunrise-III}/SUSI near-UV spectroscopy to characterise how oscillatory power reorganises along a single transect that crosses multiple magnetic and thermodynamic environments, including a sunspot (umbra and penumbra), a nearby pore, plage, and weak-field surroundings. Using line-core Doppler-velocity time series from a set of 30 relatively unblended spectral lines, we computed Morlet-based refined global wavelet spectra (RGWS) and formed band-integrated power maps in three frequency ranges (2--4\,mHz, 4--6\,mHz, and 6--12\,mHz). The resulting stacked maps provide a compact, line-resolved view of how the frequency-dependent power distribution varies laterally along the slit, while the median (plus IQR; Figure~{\ref{fig:rgws_bandpass_stats}}) ensemble summary quantifies where the redistribution is most coherent across the line set. Because each line's band power is normalised by its maximum along the slit (Section~\ref{subsec:slit_power_maps}), these maps quantify the relative redistribution of significant power between environments for a given line, rather than absolute power differences between lines or regions.

Across the line ensemble, three robust observational trends emerge that broadly align with the established magnetic structuring of oscillatory power. First, in the weak-field and plage segments the 2–4\,mHz band commonly shows strong relative power, consistent with the ubiquitous photospheric dominance of $p$-mode oscillations around five minutes and their enhanced visibility in relatively weak magnetic regions \citep[e.g.,][]{1970ApJ...162..993U, 1997ApJ...488..462R, 2011ApJ...730L..37M, Lagg+2026}. In contrast, the same 2–4\,mHz band is frequently reduced in relative strength within the strongest-field concentrations (umbra and pore cores) compared to its maxima in the weak-field/plage parts of the slit, in line with the long-established reduction of low-frequency power in sunspot umbrae \citep[e.g.,][]{2006RSPTA.364..313B, 2015LRSP...12....6K}. We emphasise that this does not imply an absence of 2--4\,mHz power in the umbra; rather, it indicates that for most diagnostics the strongest 2--4\,mHz band-integrated power along this transect occurs outside the umbral core. This is consistent with Paper~I, where line-by-line RGWS shapes in the umbral core often show substantial power in the 2--4\,mHz range even when higher-frequency components are also present. Notably, beyond the umbra/pore cores we also find a pronounced 2–4\,mHz suppression within a penumbral segment dominated by fibrillar structures (see below). Although 5-min (2--4\,mHz) power enhancements have occasionally been reported even in umbral Doppler-velocity measurements \citep[e.g.,][]{2021RSPTA.37900175N}, the apparent balance between the five- and three-minute bands may be sensitive to spatial resolution and stray-light. In particular, increased quiet-Sun stray-light contamination can systematically shift the apparent dominant peak from $\sim$5 to $\sim$3\,mHz in umbral spectra \citep{2026ApJ...997..197B}. In that context, the relative suppression of 2--4\,mHz power that we find in the umbra and pore cores is consistent with expectations for the high-resolution, low-stray-light observations analysed here.

Second, the intermediate band (4--6\,mHz; encompassing the three-minute oscillations) shows strong power enhancement in the umbral core and across the same fibril-dominated penumbral segment extending close to the umbra–penumbra boundary in most spectral lines, as well as displaying power peaks in the pore and part of the plage in several diagnostics. This is consistent with many reports that three-minute power is prominent in sunspot and pore atmospheres and can be especially apparent near umbral-penumbral boundaries \citep[e.g.,][]{2013ApJ...779..168J, 2015SSRv..190..103J, 2015LRSP...12....6K, 2022ApJ...938..143G}. Importantly, the umbral/pore three-minute signal is not purely a chromospheric phenomenon: it has been reported already at photospheric heights in both pores and sunspot umbrae using photospheric spectral lines \citep{2012A&A...539L...4S,2017ApJ...836...18C,2019ApJ...883...72C}. At the same time, statistical HMI studies indicate that photospheric umbral spectra can exhibit a strong peak near $\sim$3.3\,mHz while still showing distinct secondary power concentrations in the 4--6\,mHz band \citep{2025A&A...697A.156B}, while keeping in mind that instrumental resolution/height mixing and stray-light can influence the apparent balance between the three- and five-minute bands in photospheric umbral measurements \citep{2026ApJ...997..197B}. 

Third, the high-frequency band (6–12\,mHz) is also strongly concentrated in the umbra and pore cores, demonstrating substantial high-frequency content in strongly magnetised atmospheres. This agrees with a growing body of work reporting multi-component spectra and significant higher-frequency contributions in sunspots and pores, frequently discussed in terms of multiple MHD wave modes as well as eigenmode families whose frequency content depends on structure size and magnetic topology \citep[e.g.,][]{2021A&A...649A.169S, 2022NatCo..13..479S, 2023A&A...674A.109C, 2024A&A...688A...2J}.

We also note a small but systematic spatial offset within the umbra between the locations of maximum 4–6\,mHz and 6–12\,mHz power (Figures~\ref{fig:rgws_bandpass_maps} and \ref{fig:rgws_bandpass_stats}), indicating that the mid- and high-frequency components are not strictly co-spatial even within the umbral core. Similar spatial offsets between dominant frequency components have been reported in sunspot chromospheric oscillations \citep[e.g.,][]{2013ApJ...779..168J} and in photospheric pore oscillations \citep[e.g.,][]{2021A&A...649A.169S}. While we do not attempt to interpret this behaviour here, it is plausibly linked to local variations in magnetic/thermodynamic structure that modulate the transmission and/or diagnostic visibility of different frequency bands, and it is a clear target for follow-up once the fully reduced SUSI polarimetry becomes available.

Consistent with our findings across the three spectral bands, \citet{2021A&A...649A.169S} showed that the amplitude of five-minute oscillations in the quiet Sun is progressively reduced when approaching the boundary of a magnetic pore, whereas close to pore centre multiple higher-frequency components (up to $\sim$10\,mHz) appear in Doppler-velocity oscillations, interpreted as eigenmodes of a magnetic cylinder. Another recent multi-line study using the FRANCIS \citep{2023SoPh..298..146J} fibre-fed integral-field unit in the 580.7–597.3\,nm range likewise reported dominant LOS-velocity power in plage around $\sim$3–5\,mHz and multiple higher-frequency components in various photospheric lines within a sunspot umbra \citep{2026FrASS...Grant_inprep}.

Along our transect, the penumbral-fibril segment (around the slit position $s\!\sim\!1200\,\mathrm{px}$) is distinctive in that it combines (i) a pronounced suppression of 2--4\,mHz power and (ii) a strong enhancement in 4--6\,mHz across a large fraction of the line set. At the level of this Letter, we treat this as an empirical result rather than assigning a single physical cause. A plausible explanation is that this segment samples a particularly structured and inclined penumbral environment (and/or rapid fibril-associated dynamics) that shifts the local mix of motions toward shorter periods relative to nearby weak-field pixels. Minute-scale periodicities are widely reported in penumbral fibrillar chromospheric structures \citep[e.g.,][]{2011ApJ...739...92P, 2024ApJ...970...66B}; while our diagnostics primarily sample photospheric to low-chromospheric layers (Appendix~\ref{sec:appendix_eb}), we retain this comparison only as a qualitative timescale connection and defer a targeted investigation of the fibril segment to future work with improved magnetic and thermodynamic constraints. That the penumbral development on the opposite umbral side (around $s\!\sim\!550\,\mathrm{px}$) is relatively weaker, further suggests sensitivity to local fine structure and viewing/geometry along a 1D cut. It is also possible that nearby fine structure (e.g., the vicinity of a light-bridge environment on that side of the spot) could contribute to the asymmetry, as light bridges have been reported to host distinct oscillatory behaviour and frequency-dependent power changes \citep[e.g.,][]{2014ApJ...792...41Y}.

An important new result presented in this Letter is that even within a narrow near-UV spectral window, the inferred lateral organisation of oscillatory power is not identical from line to line. Many spectral lines agree on the broad redistribution trends between solar environments, but the detailed spatial pattern and the relative prominence of individual frequency bands vary substantially across the 30-line set. Thus, the main advance is not simply that different solar features favour different frequency ranges, but that this frequency-dependent spatial organisation is itself strongly line-dependent along the same active-region transect.
This line-to-line diversity is physically expected because different near-UV lines have different radiative sensitivities and height-weighting, and thus differ in their ``visibility'' to velocity perturbations \citep[and to thermodynamic or opacity/source-function variations that can influence fitted line-core shifts, e.g.,][]{1992ApJ...398..375R,2008A&A...480..515C,2012ApJ...749..136L}. The stacked maps and ensemble profiles show that the redistribution trends are broadly coherent across the line set while remaining line dependent; this dependence is not a subtle detail, because the inferred band distribution -- and even where along the slit power appears concentrated -- can shift with the chosen line. A practical consequence is that single-line studies may sample only a subset of the locally present oscillatory spectrum, whereas simultaneous multi-line spectroscopy, as presented here, can reveal additional components within the same magnetic environment. Disentangling which differences reflect height sensitivity, radiative-transfer effects, or genuine differences in wave-mode content will benefit from more detailed follow-up studies.

In this Letter we deliberately adopt a Stokes-$I$-only, line-resolved approach to establish robust observational trends across a strongly structured active-region transect. We provide Eddington--Barbier formation-height estimates only as contextual guidance (Appendix~\ref{sec:appendix_eb}); more quantitative interpretation will benefit from response functions to velocity, temperature, and magnetic field (and NLTE treatment where relevant), together with fully reduced/calibrated polarimetric products to constrain magnetic geometry and line-formation conditions along the transect. Such polarimetric information will be particularly valuable for interpreting the more dissimilar line-dependent patterns, especially where magnetic structuring and radiative-transfer effects may both contribute. The same multi-line dataset is also well suited to future cross-spectral phase/coherence analyses, which can more directly constrain propagation properties and possible variations of phase speed with height. These next steps will enable the observed band-dependent redistribution -- including the penumbral-fibril behaviour -- to be related more quantitatively to the local magnetic and thermodynamic structuring and to the wave-mode content of sunspot and pore atmospheres. Even at this stage, the Letter establishes a clear observational baseline: in a single, co-temporal, seeing-free dataset, near-UV velocity oscillation power is not only strongly environment-dependent but also measurably line dependent from line to line, and multi-line spectroscopy provides a qualitatively new lever for testing and refining the traditional picture of photospheric/low-chromospheric oscillations across magnetically diverse active-region scenes.

\begin{acknowledgments}
SJ and DBJ acknowledge support from the UK Science and Technology Facilities Council (STFC) through consolidated grants ST/T00021X/1 and ST/X000923/1. SJ also received support from the Rosseland Centre for Solar Physics (RoCS), University of Oslo, Norway. DBJ further acknowledges funding from the Leverhulme Trust (Research Project Grant RPG-2019-371) and from the UK Space Agency via the National Space Technology Programme (grant SSc-009).
We wish to acknowledge scientific discussions with the Waves in the Lower Solar Atmosphere (WaLSA; \href{https://WaLSA.team}{www.WaLSA.team}) team, which has been supported by the Research Council of Norway (project no. 262622), The Royal Society (award no. Hooke18b/SCTM; \citealt{2021RSPTA.37900169J}), and the International Space Science Institute (ISSI Team 502).
\textsc{Sunrise iii} is supported by funding from the Max-Planck-F\"orderstiftung (Max Planck Foundation), NASA under Grants \#80NSSC18K0934 and \#80NSSC24M0024 (``Heliophysics Low Cost Access to Space'' program), and the ISAS/JAXA Small Mission-of-Opportunity program and JSPS KAKENHI Grant Numbers JP18H05234 and JP23K25916. This research has received financial support from the European Union's Horizon 2020 research and innovation programme under grant agreement No.~824135 (SOLARNET) and No.~101097844 (WINSUN) from the European Research Council (ERC). It has also been funded by the Deutsches Zentrum f\"ur Luft- und Raumfahrt e.V.\ (DLR, grant no.~50~OO~1608). The Spanish contributions have been funded by the Spanish MCIN/AEI under projects RTI2018-096886-B-C5 and PID2021-125325OB-C5, and from ``Center of Excellence Severo Ochoa'' awards to IAA-CSIC (SEV-2017-0709, CEX2021-001131-S), all co-funded by European REDEF funds, ``A way of making Europe''.
\end{acknowledgments}

\appendix

\section{Eddington--Barbier formation-height estimates}
\label{sec:appendix_eb}

For contextual guidance only, we provide approximate formation-height estimates for the analysed near-UV spectral-line diagnostics using the Eddington--Barbier (EB) approximation, i.e., the geometric height where $\tau_\lambda=1$ for each wavelength sample \citep[e.g.,][]{2021arXiv210302369R}. We report EB heights for three representative atmospheres: the quiet-Sun FALC and plage FALP semi-empirical models \citep{1993ApJ...406..319F}, and a representative sunspot (umbral-core) atmosphere extracted from a radiative-MHD (MURaM) simulation \citep{2005A&A...429..335V,2012ApJ...750...62R}. These EB heights are included only as a qualitative reference; they are not used as a primary interpretation axis in this Letter. We note that these EB values should be regarded as representative reference heights: FALC and FALP are 1D average atmospheres, and in structured 3D atmospheres (particularly sunspots) the effective $\tau_\lambda\!=\!1$ height can vary substantially across the field of view, by several hundred kilometres, depending on local conditions.

We adopt EB here because it provides a consistent, window-wide height proxy for all wavelength samples and identified lines, enabling a compact contextual comparison between atmospheres. More complete height-sensitivity diagnostics (e.g., response functions to velocity, temperature, and magnetic field, and/or NLTE modelling) are substantially more computationally demanding and are best pursued alongside a fully consistent treatment of the atmospheric stratification and magnetic geometry. In addition, at this stage the present Letter is restricted to Stokes~$I$ products, while the full-Stokes reduction and subsequent inversion-based constraints are being finalised; those will support more rigorous response-function and NLTE assessments in follow-up work.

To compute the EB heights we performed LTE radiative-transfer calculations with the RH code \citep{2001ApJ...557..389U,2015A&A...574A...3P}, using atomic line parameters from the Kurucz line lists \citep{1995all..book.....K}. For each spectral line, we determine a representative wavelength from the model spectrum and record the corresponding geometric height where $\tau_\lambda=1$ in the given atmosphere. Heights are referenced to the $\tau_{500\,\mathrm{nm}}=1$ level of the quiet-Sun model; in the sunspot atmosphere this can yield negative values due to the Wilson depression of the $\tau\approx1$ surface in strong-field regions \citep[e.g.,][]{2003A&ARv..11..153S}.

\begin{figure}[!ht]
\centering
\includegraphics[width=\linewidth]{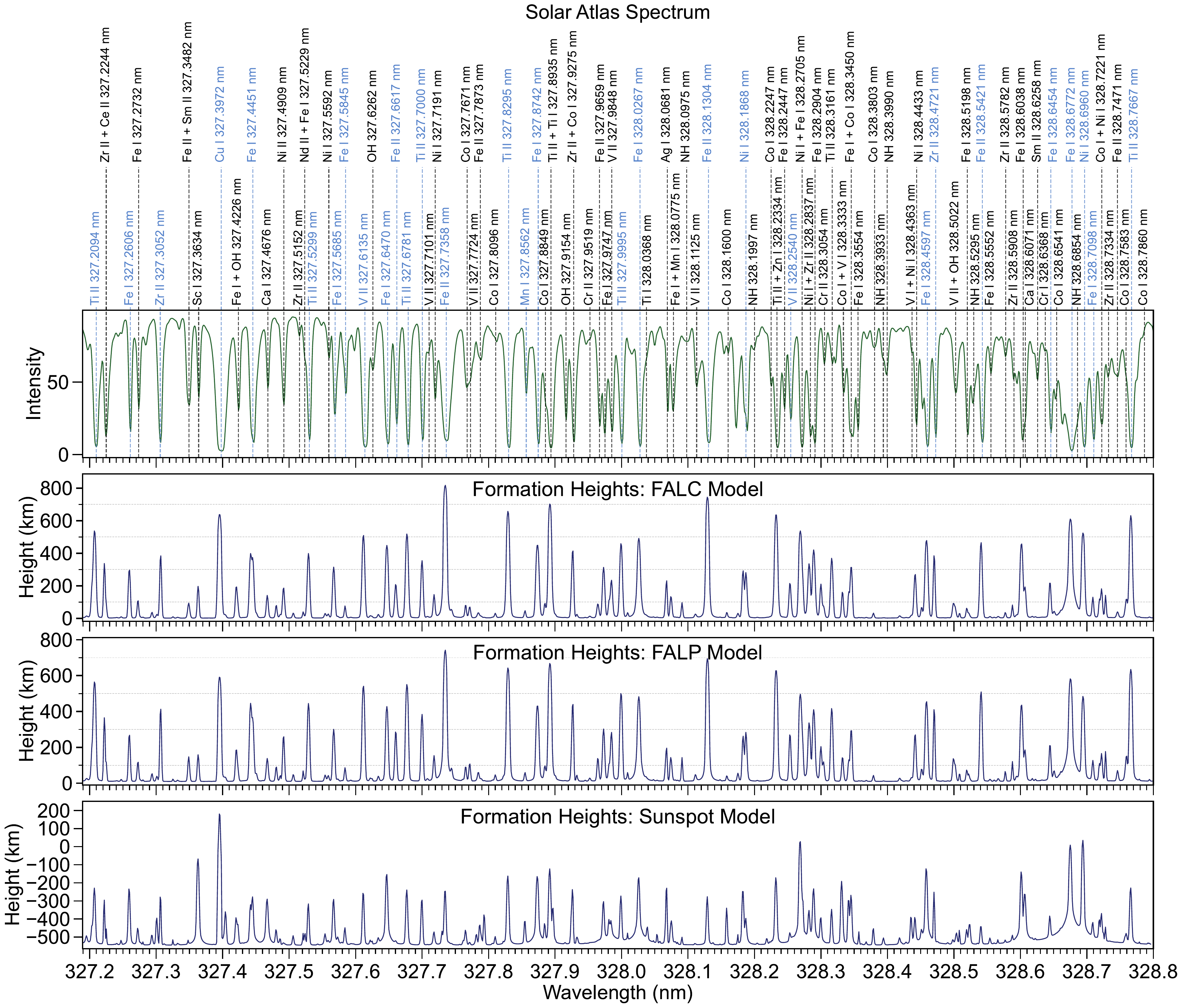}
\caption{Eddington--Barbier (EB) formation-height proxy across the 327--329\,nm window for contextual guidance. 
Top: atlas spectrum with all identified lines marked; the 30 atomic diagnostics analysed in this Letter are highlighted in blue. 
Bottom panels: EB height estimates (geometric height where $\tau_\lambda\!=\!1$ for each wavelength sample) computed in three representative atmospheres: quiet-Sun FALC, plage FALP \citep{1993ApJ...406..319F}, and a sunspot atmosphere from a radiative-MHD (MURaM) simulation \citep{2005A&A...429..335V,2012ApJ...750...62R}. EB heights are shown only as an approximate reference.}
\label{fig:eb_heights}
\end{figure}

Figure~\ref{fig:eb_heights} summarises the EB heights across the 327--329\,nm window. The top panel shows an atlas spectrum with all identified lines marked for completeness; the subset of 30 atomic diagnostics analysed in this Letter is additionally highlighted in blue. The lower panels show the corresponding EB height estimates for FALC, FALP, and the sunspot model, illustrating both the overall height range sampled within this window and the atmosphere-dependent shifts in the EB proxy.

\section{Supplementary visualisation of band-integrated power}
\label{app:rgws_additional_figures}

For completeness, we provide a qualitative Blue-Yellow-Red composite constructed from the same three row-normalised band-integrated maps shown in Figure~\ref{fig:rgws_bandpass_maps}, assigning 2--4\,mHz to blue, 4--6\,mHz to green (or yellow), and 6--12\,mHz to red. This compact view can help identify locations where the balance between low-, mid-, and high-frequency power changes rapidly along the slit. Because any such composite requires display-oriented scaling and colour mixing, we treat it as supplementary and base our interpretation on the separate band maps (Figure~\ref{fig:rgws_bandpass_maps}) and the robust ensemble profiles (Figure~\ref{fig:rgws_bandpass_stats}).

\begin{figure}[!ht]
\centering
\includegraphics[width=\linewidth]{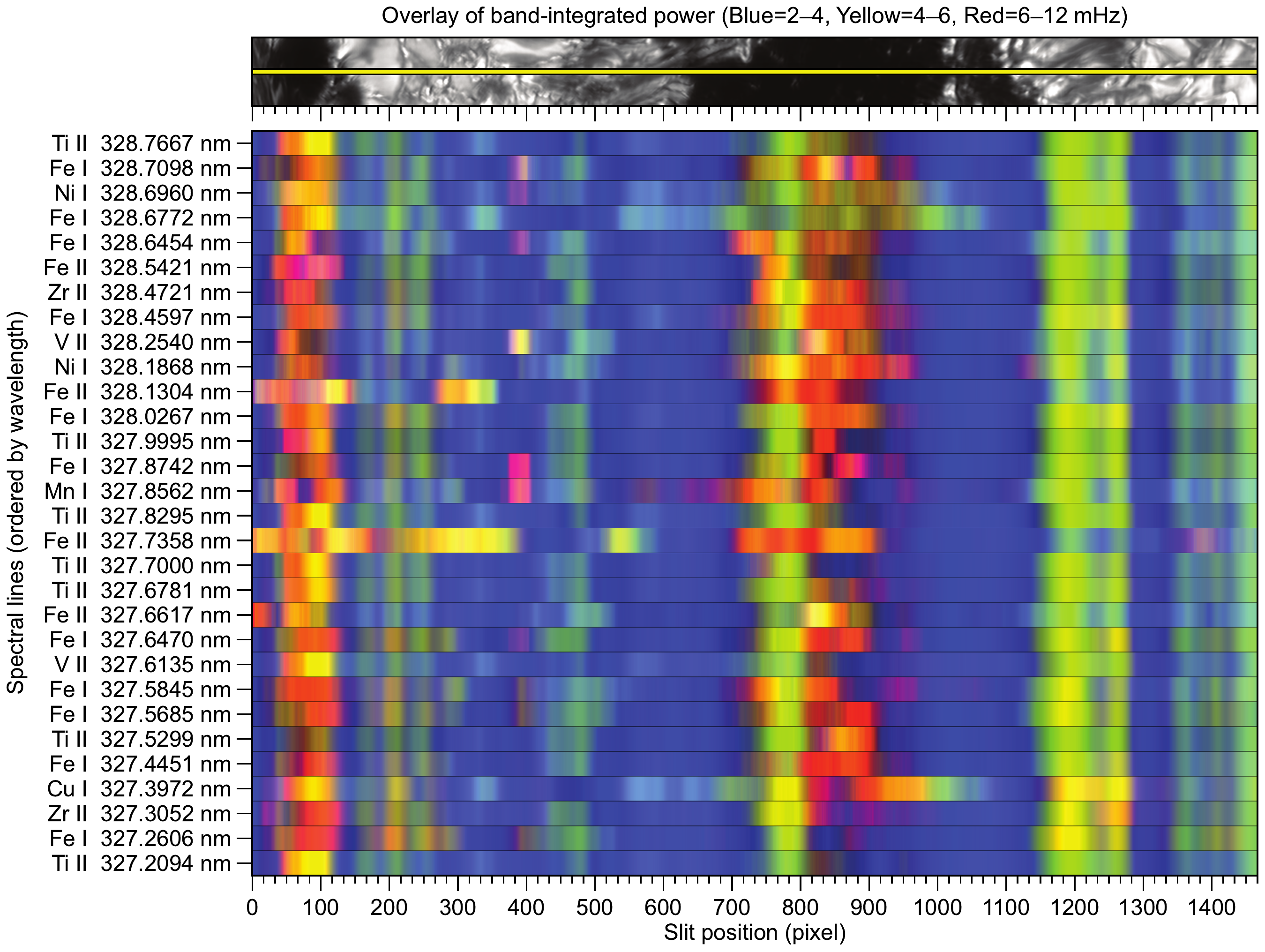}
\caption{Blue-Yellow-Red composite of the row-normalised band-integrated RGWS power along the slit, constructed from the same maps as shown in Figure~\ref{fig:rgws_bandpass_maps}. The three frequency bands are mapped to colour channels as indicated in the legend above the figure. This composite is included only as a qualitative visual summary of the relative band balance along the slit; all quantitative interpretation in the main text is based on Figures~\ref{fig:rgws_bandpass_maps} and \ref{fig:rgws_bandpass_stats}.}
\label{fig:rgws_rgb}
\end{figure}

\bibliography{article}{}
\bibliographystyle{aasjournalv7}

\end{document}